\documentclass[doublecol]{epl2}
\usepackage{amsthm}
\usepackage{amsmath}
\usepackage{amssymb}
\usepackage{color}
\usepackage{cite}
\usepackage{graphicx}
\usepackage{xcolor}
\usepackage{subfigure}
\usepackage[linktocpage]{hyperref}
\usepackage{setspace}
\usepackage{titletoc}

\newcommand{\mrd}{\mathrm{d}}

\definecolor{darkred}{rgb}{0.9,0.05,0.05}
\definecolor{darkblue}{rgb}{0.05,0.05,0.6}
\definecolor{darkgreen}{rgb}{0.05,0.6,0.05}
\definecolor{brightgreen}{rgb}{0.1,0.9,0.1}

\renewcommand*{\eqref}[1]{%
  \begingroup
    \hypersetup{
      linkcolor=darkblue,
      linkbordercolor=darkblue,
    }%
    \textcolor{darkblue}{(\ref{#1})}%
  \endgroup 
}

\hypersetup{linkcolor=red,citecolor=brightgreen,urlcolor=black,colorlinks=true}

\title{Light speed memory as a local observable for soft hairs}

\author{Kamal Hajian}

\institute{                    
Department of Physics, Middle East Technical University, 06800, Ankara, Turkey and\\
Hanse-Wissenschaftskolleg institute for advanced study, Lehmkuhlenbusch 4, 27733 
Delmenhorst, Germany
}

\abstract{
Einstein observers  in flat space-time are inertial observers which use light to synchronize their clocks. For such observers, speed of light is a constant by construction. However, one can use super-translations to change coordinates from Einstein to BMS coordinates. From the point of view of BMS observers, speed of light is not a constant all over the space-time and in all directions. So in general,  clocks which are synchronized for Einstein observers are not synchronized for BMS observers, and vice versa. Based on this fact, we propose a local observable for detecting the soft hairs, which is the variations in speed of light for such  observers.
We also investigate the relation of this observable to gravitational memory, which is a permanent change of position of test particles at infinity, after a gravitational wave passes completely from them.  It is shown that the BMS time coordinate is the physical time after a gravitational wave, and it is the legitimate time to be used to calculate the light speed. Based on this argument, the analysis predicts a permanent change in the speed of light rays which propagate in asymptotics after a gravitational wave. Moreover,  it is explained how this change is related to the gravitational memory by comparing their significance in observations. }

\begin{document}

\maketitle

\section{Introduction}
Studying diffeomorphisms as symmetries of covariant gravitational theories has been a fruitful line of research after introduction of general relativity. One of the interesting achievements in this subject has been realization of asymptotic symmetry groups (ASG). ASG is a subset of diffeomorphisms which has non-trivial conserved charges, and can carry non-trivial physical contents. It has been studied for different asymptotics and boundary conditions (see examples in \cite{BMS,Sachs,Brown:1986nw,Barnich:2009se,Carlip:1999cy,Guica:2008mu}). Especially, for the asymptotically flat space-times in $4$ dimensions, the group of asymptotic symmetries has been realized to be the BMS$_4$ group, consisted of super-translations \cite{BMS, Sachs} and super-rotations \cite{Barnich:2009se}. 

{An interesting consequence of realization of BMS$_4$ as an ASG is that it makes the vacuum of the theory to be degenerate. Different states of the vacuum  are labeled by ASG conserved charges. The presence of new conserved charges associated with such a group of symmetries has provided new insights towards resolving black hole microstate problem \cite{Bekenstein:1973ur} (\textit{e.g.} see \cite{Hawking:2016msc}), as well as information paradox \cite{Hawking:1976ra} (\textit{e.g.} as in \cite{Hawking:2015qqa}). In this regard, the term ``soft hair" was coined by Hawking-Perry-Strominger for these charges \cite{Hawking:2016msc}, motivated by the terminology of black hole No-Hair theorem. Nonetheless, the proposal of soft hairs for black holes has been very controversial, and there have been challenging arguments in its opposition (see \textit{e.g.} \cite{Mirbabayi:2016axw,Bousso:2017dny,Bousso:2017rsx,Javadinezhad:2018urv,Compere:2019rof}).} 

The observability of the ASG charges, the soft charges, are under question too. So far, to the best of our knowledge, there is not yet any proposal for direct local observation of soft hairs and soft charges (find reviews in \textit{e.g.} \cite{Brown:1986nw} for canonical and \cite{Hajian:2015xlp} for covariant calculations of soft charges).  However, there is an interesting indirect method to detect the soft hairs, which is called gravitational memory (GM) \cite{Braginsky:1986ia,Christodoulou:1991cr,Thorne:1992sdb}. GM is a permanent displacement in the position of two test particles at infinity, when a gravitational wave (GW) propagates and passes from them. It has been shown that a GW can change the asymptotic geometry, and variate the soft hairs and soft charges. The connection between these two phenomena, the GM and changes in soft charges of ASG, has been discovered by Strominger and Zhiboedov in \cite{Strominger:2014pwa}.   

{In this paper, we suggest another local observable to detect the soft hairs. Before delving into the proposal, it is important to clarify what we mean by the term ``soft hair," because in the literature it has been used interchangeably for ASG gauge generators, their charges, and action of them on the dynamical fields. By ``soft hairs" in this paper we mean the latter description, i.e. we focus on different metrics which are related to each other by the action of ASG (see \textit{e.g.} section 7.2 in Ref. \cite{Strominger:2017zoo} for more details).   The proposal to distinguish such field configurations is based on detection of speed of light rays which propagate on these backgrounds.} The main point here, which is conceptually the cornerstone of this paper, is that the speed of light $c$ is the norm of a 3-vector velocity of light, which is not an invariant quantity, and can be different for different observers. To clarify this, we remind a simple example. Let us conventionally choose the speed of light in asymptotic flat region to be $1$. Then, the speed of a light ray which approaches radially to a Schwarzschild black hole horizon tends to vanish,  $c\to 0$ for all observers standing outside the black hole, while, for observers free-falling around the horizon, $c=1$.  This is a standard example showing the observer-dependency of the speed of light. Having this in our mind, the idea behind the analysis in this paper is based on searching for a natural  observer-dependent quantity to distinguish BMS observers, which we found to be the speed of light.

The paper is organized as follows. In the next section, we revisit Einstein observers whose speed of light is conventionally chosen to be a constant. Then, the BMS observers are briefly reviewed, and it is shown that for them the speed of light is generally not a constant. In section 3, the GM is reviewed, and finally in section 4, we connect the results in section 2 and 3, i.e. the light speed and GM. In the last section, it will be shown that the BMS time coordinate is the physical time by which the speed of light should be measured after a GW.

\section{Einstein observers vs. BMS observers}\label{section Ein}
\noindent\textbf{Einstein observers:} The 4 dimensional Minkowski space-time in Cartesian coordinates is simply
\begin{equation}\label{Ein metric}
\mrd s^2=-\mrd t^2+\mrd x_1^2+\mrd x_2^2+\mrd x_3^2,
\end{equation}
in which $t$ and $\vec{x}=(x_1,x_2,x_3)$ denote time and space coordinates respectively. The signature of the metric is chosen to be $(-,+,+,+)$. Observers in these coordinates, which we call \emph{Einstein observers}, measure the speed of light  to be equal to $1$ everywhere in the space-time and in any direction. This terminology originates from Einstein synchronization method of clocks. In his method, clocks (which are at rest in different points of space) are synchronized using light rays.  Each clock is set such that the speed of light would be a constant when it is measured at any position and moment, and in any spatial direction. Let us give a simple example. Consider two clocks at rest on the $x_1$ axis, with the spatial distance $\Delta x_1=\ell$. One of the clocks is set to zero, while sending a light ray from it towards the other clock. When the light reaches the other clock, that clock is set to $\ell$.  This is the method which Einstein used to define simultaneity in all rest frames in special relativity. Nonetheless, in general such a synchronization is conventional, and one can use another method to define simultaneity.    

In ASG analysis at null infinity of asymptotic-flat space-times, it is conventional to use $x^\mu=(u,r,z,\bar z)$ coordinates instead of Cartesian coordinates via  transformations
\begin{align}
r^2=(\vec{x})^2, \, \, u=t-r, \, \, z=\frac{x_1+ix_2}{x_3+r}, \, \, \bar z=\frac{x_1-ix_2}{x_3+r}
\end{align} 
to re-write the Minkowski metric as 
\begin{equation}\label{ASG metric}
\mrd s^2=-\mrd u^2-2\mrd u \mrd r +2 r^2\gamma_{z\bar z} \mrd z \mrd \bar z, \, \, \gamma_{z \bar z}=\frac{2}{(1+z\bar z)^2}.
\end{equation}
In this coordinate, $z$ runs over the whole complex plane, north pole is at $z=0$, equator is at $z\bar  z=1$, and south pole is at $z\to \infty$. It is easy to check that in this coordinate, the speed of light is kept intact, and is equal to the constant $1$. Let us check this for radial light rays, as well as tangential light rays on equator. For the radial light rays $\mrd z=\mrd \bar z=0$. The speed of light is defined via derivative of radial length element with respect to time, which is ${\mrd r}/{\mrd t}$.  The relation $\mrd u = \mrd t - \mrd r$ and the null condition $\mrd s^2=0$ in the metric \eqref{ASG metric} yields ${\mrd r}/{\mrd t}=1$, which is the expected speed of light.  For the tangential light rays on equator, $z\bar z=1$ and $\mrd r=0$. Accordingly, $\mrd u = \mrd t$, and so the $u$ coordinate can be used as the time coordinate in calculation of light speed.  Besides, on the equator $\gamma_{z \bar z}=1/2$, and the spatial line element is equal to $ \sqrt{r^2 \mrd z \mrd \bar z}$. Hence, the speed of light is read to be $ \sqrt{r^2 \mrd z \mrd \bar z}/\mrd u$.  Requesting $\mrd s^2=0$ in the metric \eqref{ASG metric}, provides 
\begin{equation}\label{c Ein}
c=\frac{\sqrt{r^2 \mrd z \mrd \bar z}}{\mrd u}=1.
\end{equation}

\noindent\textbf{BMS observers:} In order to distinguish the BMS observers, it is useful to introduce a notation. If we denote the coordinates $(z,\bar z)$ by the Latin indices $a,b,c, \ldots$, then $D_a$ means a covariant derivative on the 2-spheres of constant $u$ and $r$. In other words, one uses the $\gamma_{ab}$ matrix (with components $\gamma_{\bar z z}= \gamma_{z \bar z}$ in \eqref{ASG metric}, and $\gamma_{zz}=\gamma_{\bar z\bar z}=0$) to define covariant derivative on the spheres. In addition, the inverse of matrix $\gamma_{ab}$, which is denoted by $\gamma^{ab}$, is used to raise the Latin indices.  

Now we are ready to apply super-translations on the Einstein coordinates $x^\mu$ and metric in \eqref{ASG metric}, to introduce the BMS observers. A generic super-translation is generated by the following vector fields \cite{BMS,Sachs}:
\begin{equation}\label{BMS generator}
\zeta[f]=f\partial_u-\frac{1}{r}(D^z f \partial_z + D^{\bar z}f \partial_{\bar z})+ D^z D_z f \partial_r + \cdots ,
\end{equation} 
in which $f$ is a function of coordinates on the sphere, i.e. $f(z,\bar z)$. The ``$\cdots$" denotes extra terms that are sub-leading in orders in $r$, which are not important in the analysis in this paper.  Applying such an infinitesimal  transformation on the coordinates $x^\mu$ and the metric  \eqref{ASG metric} results in the new coordinates $x'^\mu=x^\mu+\zeta^\mu$ and metric below\footnote{For simplicity, we keep our discussion around the Minkowski space-time in Einstein coordinates. In other words, we only focus on $\delta C_{zz}$ and $\delta C_{\bar z \bar z}$, not the $C_{zz}$ and $C_{\bar z \bar z}$ themselves. Equivalently, one can use the GM analysis in the literature and simply put $C_{zz}=C_{\bar z \bar z}=0$. However, this simplification is just fixing the clock synchronization convention \emph{before} the GW to be the Einstein convention. The analysis in this paper is independent of this convention, as it should be.}  via $g'_{\mu\nu}=g_{\mu\nu}-\mathcal{L}_{_\zeta} g_{\mu\nu}$:
\begin{align}\label{BMS observer}
\mrd s^2=&-\mrd u'^2-2\mrd u' \mrd r' +2 r'^2\gamma_{z\bar z} \mrd z' \mrd \bar z' \nonumber \\
&+D^z \delta C_{zz}\mrd u'\mrd z' + D^{\bar z} \delta C_{\bar z \bar z}\mrd u'\mrd \bar z'\nonumber\\
&+r' \delta C_{zz}\mrd z'^2+ r' \delta C_{\bar z\bar z}\mrd \bar z'^2 + \cdots.
\end{align}
We have denoted the new coordinates with prime to  carefully distinguish the new family of  observers associated with them. The observers associated with the new coordinates $x'^\mu$ are called \emph{BMS observers}. The new functions in the new metric \eqref{BMS observer}, the BMS metric, are related to the $f(z,\bar z)$ by the following constraints:
\begin{align}
&\delta C_{ab}\equiv \begin{pmatrix}
\delta C_{zz} & 0 \\
0 & \delta C_{\bar z \bar z}
\end{pmatrix}, \nonumber\\
&\delta C_{zz}=2D_z D_z f, \qquad 
 \delta C_{\bar z \bar z}=2D_{\bar z}D_{\bar z} f.
\end{align}

\noindent\textbf{Speed of light:} The light speed in BMS coordinates is not a constant in all points of the space-time, and in all spatial directions. This could be guessed by noting that the Poincar\'e group consists in the largest (linear) transformations which keep the light speed invariant in flat space-time.  The BMS super-translation as a group  is much larger group than the translations in Poincar\'e group, and one could expect that they change the light speed.   Nonetheless, one should check this guess explicitly, because the BMS transformations are non-linear transformations. It suffices to show that at some points of space-time and in some directions in BMS coordinates, the light speed is not equal to $1$. To this end, let us consider the tangential light rays on the $z\bar z=1$. For tangential motions $\mrd r'=0$, and so $u'$ is an appropriate time coordinate to calculate the light speed. The spatial line element which is tangent to the spheres of constant $u'$ and $r'$ can be read from the metric \eqref{BMS observer} by considering  $\mrd u'=0$ and $\mrd r'=0$, which yields $ \sqrt{r'^2 \mrd z' \mrd \bar z' + r' \delta C_{zz}\mrd z'^2+ r' \delta C_{\bar z\bar z}\mrd \bar z'^2}$. Imposing the null condition $\mrd s^2=0$ in BMS metric, the BMS tangential speed of light is read as 
\begin{align}
c_{_\text{BMS}}&\equiv \frac{\sqrt{r'^2 \mrd z' \mrd \bar z' + r' \delta C_{zz}\mrd z'^2+ r' \delta C_{\bar z\bar z}\mrd \bar z'^2}}{\mrd u'}\\
&=\sqrt{1-D^z \delta C_{zz}\frac{\mrd z'}{\mrd u'}-D^{\bar z} \delta C_{\bar z \bar z}\frac{\mrd \bar z'}{\mrd u'}}, \label{c BMS}
\end{align}    
which generally differs from $1$. This result explicitly shows that  Einstein and BMS observers detect different speeds of light. Moreover, the clocks which are synchronized for Einstein observers, are not synchronized for BMS observers, and vice versa. 

It is worth emphasizing that having $c\neq 1$ is  not in contradiction with invariance of the light cone for all observers, and the fact that the light moves on the light cone. Actually, we have used this fact explicitly when we put $ds^2=0$ to derive \eqref{c BMS}. 

Notice also that Einstein and BMS observers not only differ in synchronization of their clocks, but also in time+space decomposition and labeling their positions. This can be seen directly from the BMS generators in \eqref{BMS generator}, which change other coordinates in addition to the time $u$. One also might be concerned about how it is possible to distinguish $u'$ as a physical time to calculate the light speed. This issue will be addressed in the last section.

\section{Gravitational memory for BMS observers}

In this section, GM analysis in the language of Strominger and Zhiboedov in \cite{Strominger:2014pwa} (which is pedagogically presented in \cite{Strominger:2017zoo}) is reviewed. For book keeping, we will not repeat all the details of calculations.  An interested reader can find the details in the references. However, the analysis is reported such that the role of BMS observers and their coordinates would be explicit.

In the set up proposed for observing gravitational memory, one begins with an asymptotically flat background metric. Using the convention in clock synchronization,  the initial metric can asymptotically chosen to be in the form of Einstein metric \eqref{Ein metric}. In this background, one places two test particles (or detectors) at rest, labeled by numbers $1$ and $2$, in the spatial coordinates  $\vec X_1=(r_0, z_0, \bar z_0)$ and  $\vec X_2=(r_0, z_0+\Delta z, \bar z_0+\Delta \bar z)$ respectively. It is assumed that $r_0$, which is the common radius of the test particles, is very large compared to the scales of the matter contents in the bulk.   In other words, the test particles are installed in asymptotics. Moreover, it is also assumed that $\Delta z$ and $\Delta \bar z$ are infinitesimal and of order $\sim 1/r_0 $, to provide finite spatial distances in the limit $r_0\to \infty$. Over time, each one of the test particles constitute a world line in the space-time. For example,  if the particles are not forced to change their positions, their world lines are identified by $X_1^\mu=(u, r_0, z_0, \bar z_0)$ and   $X_2^\mu=(u, r_0, z_0+\Delta z, \bar z_0+\Delta \bar z)$ in which $u$ is a variable.    

A covariant distance is by definition a relation between two events, not two world lines. So in general, especially when the system is not static (as is true in the case when a GW passes), one cannot define a covariant distance between two world lines.  But, it is possible to define spatial distances, when the system reaches to a static configuration. In other words, depending on the choice of a system of coordinates, one defines the spatial coordinate difference $\Delta X^\mu$ and spatial distance $\Delta s$ as
\begin{equation}\label{distance}
\Delta X^\mu\equiv (X_2^\mu-X_1^\mu)\Big|_{u_\ast}, \qquad \Delta s\equiv \sqrt{\Delta X_\mu \Delta X^\mu}.
\end{equation}
The time which is denoted by $u_\ast$ identifies the hyper-surface of simultaneity at which the measurements are performed. It is worth emphasizing that although the terms in \eqref{distance} look like to be covariant,  they depend on the choice of coordinates, and hence are sensitive to the choice of  the observers. 

\noindent\textbf{Gravitational memory:} By preparing the setup above, and fixing the clock synchronization convention to be \textit{e.g.} the Einstein convention, let a burst of matter, or two coalescing black holes, or any other possible source radiate a pulse of GW,  propagating  towards the null infinity. The time when the pulse reaches the test particles at radius $r_0$ is denoted by $u_0$, and the time when it passes completely from the particles is denoted by $u_0+\delta u$.  Before and at the time $u_0$, the metric is (conventionally chosen to be) the Einstein metric \eqref{ASG metric} (which will be denoted by $g_{\mu\nu}$, i.e. without prime). In addition, the spatial distance (as measured by Einstein observers) between the test particles is   
\begin{align}
&\Delta X^\mu= (X_2^\mu-X_1^\mu)\Big|_{u_0}=(0,0,\Delta z, \Delta \bar z), \\
 &\Delta s=\sqrt{2r_0^2\gamma_{z_0 \bar z_0}\Delta z \Delta \bar z}\equiv L_0. \label{del X}
\end{align}

The analysis of dynamics of such a setup in the literature  shows that after the GW passes, the metric changes from the Einstein metric \eqref{ASG metric} to BMS metric \eqref{BMS observer}. Besides, the $\Delta s$ changes by a non-zero term. These two changes are related to each other. This is the connection between ASG (incarnated in BMS metrics) and GM (saved in changes in particles distance) which was discovered by Strominger and Zhiboedov in 2014 \cite{Strominger:2014pwa}.  To make this paper self-contained, we repeat the main steps in their analysis. However to be clear and as simple as possible,  the details are removed and the final results are simply reported  (see exercise 13 and answer to it in \cite{Strominger:2017zoo} to find all details). Let us denote the metric and world line of particles after the GW by primes, i.e. as $g'_{\mu\nu}$ and $X'^\mu$. 
\begin{enumerate}
\item In order to calculate $\delta \Delta s$, one needs to calculate $g'_{\mu\nu}=g_{\mu\nu}+\delta g_{\mu\nu}$ and $\Delta X'^\mu=\Delta X^\mu+\delta \Delta X^\mu$.  After finding these quantities, they are inserted in the following equation
\begin{equation}\label{del Del s}
\delta \Delta s  \equiv \sqrt{ g'_{\mu\nu}\Delta X'^\mu\Delta X'^\nu}-\sqrt{ g_{\mu\nu} \Delta X^\mu \Delta X^\nu},
\end{equation}
while keeping to the appropriate powers of  $\delta$ and orders of $r_0$.  
\item To find $ g'_{\mu\nu}$, the dynamics of the metric is analyzed using the Einstein field equations, when a GW passes from initial time $u_0$ to a final time $u_0+\delta u$. The result turns out to be the  BMS metric in \eqref{BMS observer}.  The functions $\delta C_{zz}$ and $\delta C_{\bar z \bar z}$ can be calculated from specifications of the GW. However the explicit functionality of these functions is not important in our discussion.
\item The $ \Delta X'^\mu$ is found by studying the geodesics of each particles. To this end, one needs to solve the geodesic equation $v^\mu\nabla_\mu v^\nu=0$, with $v_0^\mu=(1, 0, 0, 0)$ as initial velocity for each one of the particles, and  $X_{0,1}^\mu = (u_0, r_0,z_0, \bar z_0)$ and $X^\mu _{0,2}=(u_0, r_0,z_0+\Delta z, \bar z_0+\Delta \bar z)$ as the initial positions of the particles $1$ and $2$ respectively. The bottom-line of the calculation, to the relevant order of $1/r$ expansion, is as follows:
\begin{itemize}
\item the velocities $v^\mu$ for each one of particles approximately do not change during the evolution, i.e. $v'^\mu \approx v^\mu_0$,
\item $\Delta X^\mu$ does not change either. Expressing carefully, 
\begin{equation}\label{BMS delX}
\Delta X'^\mu \equiv  (X_2^\mu-X_1^\mu)\Big|_{u_0+\delta u} \approx (0,0,\Delta z, \Delta \bar z).
\end{equation}
\end{itemize}
Notice that we have dropped the terms  in the results above  which in the expansions of $\frac{1}{r_0}$ would be irrelevant to the GM, i.e. these results are reported such that eventually they are enough to find the desired results, which is the leading variation in $\Delta s$.
\item The final step is calculating $\delta \Delta s$ using the results derived above. Inserting the $g'_{\mu\nu}$ from \eqref{BMS observer} and $\Delta X'^\mu$ from \eqref{BMS delX} into the \eqref{del Del s}, then
\begin{align}
\delta \Delta s  &=\sqrt{L_0^2 + r_0 \delta C_{zz} \Delta z^2 +  r_0 \delta C_{\bar z  \bar z} \Delta \bar z^2}-L_0\\
&\approx \frac{r_0}{2L_0}(\delta C_{zz} \Delta z^2 +  \delta C_{\bar z  \bar z} \Delta \bar z^2). \label{GM}
\end{align}
\end{enumerate}
This is the final result in the GM analysis. From this result it can be concluded that the functions $\delta C_{zz}$ and $\delta C_{\bar z\bar z}$ which play a major role in calculating charges for the super-translations, can be detected by the GM.

By the notation $u'=u_0+\delta u$,  it becomes clear from \eqref{BMS delX} that in the derivation of the result above, one has used the surfaces of constant time in BMS coordinates. It will be explained  in the next section how the light speed, which is also observer dependent, can be used as a local observable to measure the soft changes in the space-time after a GW.

\section{Gravitational memory is saved in speed of light}
Let us begin with observers at rest in the asymptotic Minkowski space-time. The observers have used the convention of clock synchronization to put the speed of light equal to $1$, before a GW arrives. So, in the terminology and notation used in the previous sections, the observers are Einstein observers, and
\begin{equation}
c=1\qquad\quad \text{for}\,\, u<u_0.
\end{equation}
So, the metric is the Einstein metric \eqref{ASG metric}. When a GW passes completely, the metric would be deformed to be BMS metric \eqref{BMS observer}, whose speed of light is measured to be different than $1$ in some points and directions. Especially according to \eqref{c BMS}, for the tangential speed of light on the equator
\begin{align} 
\delta c&=\sqrt{1-D^z \delta C_{zz}\frac{\mrd z'}{\mrd u'}-D^{\bar z} \delta C_{\bar z \bar z}\frac{\mrd \bar z'}{\mrd u'}}-1\label{del c exact}\\
&\approx -\frac{1}{2}(D^z \delta C_{zz}\frac{\mrd z'}{\mrd u'}+D^{\bar z} \delta C_{\bar z \bar z}\frac{\mrd \bar z'}{\mrd u'}),  \quad \text{for}\,\, u>u_0+\delta u. \label{del c}
\end{align} 
This result clearly shows that after a GW the light speed will be different in different tangential directions. The functions $\delta C_{zz}$ and $\delta C_{\bar z \bar z}$ can be read from the specifications of the passed GW. This is a standard result in the literature, and here the final result is reported. Denoting the energy-momentum tensor of the GW by $T_{\mu\nu}$, then the function  $\delta C_{zz}$ can be calculated \cite{Strominger:2017zoo} by 
\begin{align}
&\delta C_{zz}(z,\bar z)=2\int \mrd ^2\hat z \,\, \gamma_{\hat z \hat {\bar z}}\,D^2_z G (z,\bar z ; \hat z , \hat{\bar z}) \times  \nonumber\\
&\Big[ \int_{u_0}^{u_0+\delta u}\mrd u \Big( T_{uu}(u,\hat z, \hat{\bar z})-\frac{1}{4\pi}\int \mrd^2\tilde z \,\,\gamma_{\tilde z\tilde{\bar z}}T_{uu}(u,\tilde z, \tilde{\bar z}) \Big) \Big], 
\end{align}
where $G$ is the Green's function satisfying
\begin{equation}
D^2_z D^2_{\bar z} G (z,\bar z ; \hat z , \hat{\bar z})=-\gamma_{z\bar z} \delta^2(z-\hat z).
\end{equation}

We will not analyze different sources of GW which lead to different $T_{\mu\nu}$ in the equation above. However, a hand-waving argument can be provided to compare the significance of the change of light speed w.r.t the significance of the GM in the test particles. To this end, from equation \eqref{del X} the approximation $\Delta z\sim \Delta \bar z \approx L_0/r_0$ can be read. Considering this  approximation in GM for test particles in equation \eqref{GM} yields 
\begin{equation}\label{GM sign}
\frac{\delta \Delta s}{\Delta s}=\frac{\delta \Delta s}{L_0}\approx \frac{1}{2r_0}(\delta C_{zz} +  \delta C_{\bar z  \bar z}).
\end{equation}
On the other hand, from the equation \eqref{c Ein} one can find the approximation $\mrd z'/\mrd u' \sim \mrd \tilde z'/\mrd u' \approx 1/r_0$. Inserting this approximation in variation of tangential light speed in \eqref{del c} results to
\begin{equation}\label{c sign}
\frac{\delta c}{c}\approx -\frac{1}{2r_0}(D^z \delta C_{zz}+D^{\bar z} \delta C_{\bar z \bar z}).
\end{equation}
It is clear from the results above that the fall-off behavior of the GM in the test particle system and the variation of light speed are similar, i.e. of order of $1/r_0$. Nonetheless, GM in the test particle system is encoded in $\delta C_{zz}$ and $\delta C_{\bar z  \bar z}$, while GM in the light speed is saved in their tangential derivatives. 

\section{Change of light speed or clock desynchronization?}
The proposal of detecting speed of light as a local measurable observable for distinguishing soft hairs (i.e. different BMS metrics) is more general than the GW setup (notice that the result in \eqref{c BMS} is a generic result compared to the GW setup). It can be used for different contexts where soft hairs exist, \textit{e.g.} in black hole microstate studies via ASG. However, focusing on the GW experiment, an acute reader might have posed a question why the coordinate $u'$ is chosen in the definition of  light speed after a GW. In other words, why $c'$ is defined by taking derivative of the line element w.r.t $u'$. This question is legitimate, because $u'$ seems to be a time \emph{coordinate}, which might be different from the \emph{physical} time. According to this ambiguity of surfaces of constant time, one could also be skeptic about the spatial line element in the definition of $c'$. However, this latter is not a real challenge, because the clocks are at rest after a GW, and the definition of spatial line element is independent of the way that clocks are synchronized. This is a fact which is at the heart of Einstein's method of synchronization, which can be explained as ``the spatial distance of clocks at rest are well-defined without identifying any clock synchronization." 

Following the question above, and interestingly, we found that the right-hand-side of the equation \eqref{del c} has also appeared in a different equation in Strominger and Zhiboedov  paper \cite{Strominger:2014pwa} (see eq.(4.9) in that reference.). For the ease of the reader, we bring their analysis in brief here, and explain why their specific argument is in contrast with the results in this paper. 

Strominger and Zhiboedov in their derivation considered that a GW has passed, and the system in the BMS metric \eqref{BMS observer}. They also considered two clocks at rest at the same radius $r_0$, in coordinates which differ by $\Delta z$ and $\Delta \bar z$. Denoting the time needed for a flash of light to travel from the first clock to the second by $\delta_{_{12}}u$, then inserting $dr=0$ and $ds^2=0$ in the BMS metric  \eqref{BMS observer} yields 
\begin{align}
r_0^2\gamma_{z\bar z}\Delta z\Delta\bar z & +r_0 \delta C_{zz}\Delta z\Delta z + D^z\delta C_{zz}\delta_{_{12}}u\, \Delta z \nonumber\\
&-\frac{1}{2}(\delta_{_{12}}u)^2+\mathrm{c.c}=0.
\end{align}
For the time needed for the light to travel in the reverse direction, i.e. from the second clock to the first one, one can change $\Delta z\to -\Delta z$ and  $\Delta \bar z\to -\Delta \bar z$ in the result above, 
\begin{align}
r_0^2\gamma_{z\bar z}\Delta z\Delta\bar z & +r_0 \delta C_{zz}\Delta z\Delta z - D^z\delta C_{zz}\delta_{_{21}}u\, \Delta z\nonumber\\
&  -\frac{1}{2}(\delta_{_{21}}u)^2+\mathrm{c.c}=0.
\end{align}
Comparing the two results above, they found  \cite{Strominger:2014pwa}
\begin{align}\label{desynch}
\delta_{_{12}}u-\delta_{_{21}}u=D^z\delta C_{zz} \Delta z+D^{\bar z}\delta C_{\bar z \bar z}\Delta \bar z.
\end{align}
Then, they assume that the  light speed to be $1$, and \emph{based on this assumption}, they deduce that  the clocks are desynchronized. Notice that the right hand side is exactly equal to our result in \eqref{del c}, with an extra factor of $2$ which is a result of the reciprocating journey of light.

In the previous sections, we showed that $c=1$ is specific to Einstein observers, not the BMS observers. Hence by assuming $c=1$, what Strominger and Zhiboedov found is ``desynchronization with respect to Einstein clocks." This last section of our paper here is provided to show that the physical clocks after a GW are BMS clocks, not Einstein clocks (if one has synchronized the physical clocks before GW to be Einstein clocks).  Hence, we show that  BMS clocks, which are  \emph{synchronized} in the coordinate $u'$, are physical. So, instead of desynchronization of clocks,   speed of light deviates from $1$, undermining the desynchronization argument.   The cornerstone of physical realization of synchronized clocks is the following proposition which is based on what can be observed in experiments:

\emph{If  clocks at rest are synchronized  by some method (\textit{e.g.} Einstein method) before a GW, then after the GW there is a privileged surface of constant time {(up to the Lorentz transformations)}, which is realized by what the clocks are indicating late enough after the GW.}

Here we show that these physically privileged surfaces of constant time are the surfaces of constant $u'$, which admits the validity of the \eqref{del c}, and rejects the (physical) clock desynchonization interpretation of  \eqref{desynch}. To this end, one can look at  the velocity 4-vectors of test particles, which can also be considered as clocks  at rest before and after a GW. Explicit calculation shows (see eq.9.0.193 in \cite{Strominger:2017zoo}) that their velocity at all times, even during the passage of a GW is approximately 
\begin{equation}
v^\mu(\lambda)\approx v_0^\mu=(1, 0, 0, 0),
\end{equation} 
where $\lambda$ parametrizes the evolution, and $\lambda=0$  is  beginning of passage of the GW. From this result, one finds that  coordinates of the clock during the evolution is approximately equal to \cite{Strominger:2017zoo} 
\begin{equation}
X^{\mu}(\lambda)\approx (u_0+\lambda, r_0, z_0, \bar z_0).
\end{equation}
However, what a clock indicates is its proper time, which we denote by $\tau$. In coordinates which the clock is at rest, the proper time at any $\lambda$  is equal to
\begin{equation}
\tau(\lambda)=\int_{\lambda=0}^\lambda \sqrt{-g_{00}(\hat\lambda)\, \, \mrd X^{0}(\hat\lambda)\,\, \mrd X^{0}(\hat\lambda)}.
\end{equation}
The $g_{00}$ is not deformed during the evolution from the Einstein metric \eqref{ASG metric} to the BMS metric \eqref{BMS observer}, and remains equal to $-1$. So, by $\mrd X^{0}(\hat\lambda)=\mrd \hat \lambda$,  the proper time is equal to 
\begin{equation}
\tau(\lambda)=\int_{\lambda=0}^\lambda \mrd \hat\lambda=\lambda.
\end{equation}
This result shows that surfaces of constant $\tau$ coincide with the surfaces of constant $X^0=u_0+\lambda$, which at the end of the evolution is the $u'$ coordinate. Therefore, physical clocks which are synchronized by the Einstein method before a GW will indicated the time in the BMS coordinate $u'$. Hence, there would not be a desynchronization of clocks at rest with respect to BMS time, in support of deviation of light speed in \eqref{del c}.

{We note that the ``privileged surfaces of constant time" in our analysis do not break the Lorentz invariance. The constant time surfaces  $u'=\text{const.}$ are ``privileged" after the GW because (conventionally) we have made the Einstein frames privileged before the GW.  Nonetheless, the Einstein frames enjoy the freedom of the Lorentz symmetry. If one applies \textit{e.g.} a boost on the initially chosen frame, then the final constant-time surfaces are boosted too, because of the Lorentz invariance of the dynamics of GWs. To show that the light speed memory in Eq. \eqref{del c} is invariant under Lorentz transformations, we note that the Einstein metric in \eqref{ASG metric} is invariant under the Lorentz Killing vectors \cite{Strominger:2017zoo}:
\begin{equation}
\zeta_{_Y}=Y^z\partial_z+\frac{u}{2}D_zY^z\partial_u+\mathrm{c.c.},
\end{equation}
in which $(Y^z,Y^{\bar z})$ is a two dimensional vector field on the $u=\text{const.}$ and $r=\text{const.}$ spheres and
\begin{equation}
Y^z=1,z,z^2,i,iz,iz^2.
\end{equation}
The final metric after the GW is related to the Einstein metric via a diffeomorphism. On the other hand, the Killing condition is invariant under the action of all diffeomorphism generators, including the BMS transformations. So, the vectors $\zeta_{_Y}$ are mapped to some $\zeta'_{_Y}$ which are Lorentz symmetries of the BMS metric in Eq.\eqref{BMS observer}. As a result, the Killing vectors $\zeta'_{_Y}$ do not change the components of the metric  \eqref{BMS observer} (which are the only inputs in the derivation of \eqref{del c}). Therefore, the light speed memory is invariant under the Lorentz Killing vectors $\zeta'_{_Y}$. 
}

\section{Conclusion}
In this paper, it was shown that the speed of light changes, when a GW passes from a distant region. The relation which quantifies this change is given in \eqref{del c}. To justify this relation, it was argued that the physical time coordinate after the GW is the BMS time coordinate. Some approximations were also provided to compare significance of changes of light speed with the standard GM via test particles. The results of the approximations are presented in \eqref{GM sign} and \eqref{c sign}. They show that both of the mentioned observables fade similarly in terms of the distance from the source of a GW. 

{Soft hairs in ASG analysis are important theoretical predictions, which can appear not only after a GW, but also in other physical setups like black hole physics.  The change of speed of light can be an appropriate observable for detecting soft hairs. The hope is that having such a sophisticated observable can open new doors to our understanding of experimental and theoretical aspects of these interesting subjects. }

\acknowledgments I am grateful for helpful discussions with Zahra Mirzaiyan, Mehdi Sadeghi, Ali Seraj, Shahin Sheikh-Jabbari and Bayram Tekin. I would like to thank Jutta Kunz, Olaf Lechtenfeld, and Mehrdad Mirbabayi for providing visiting opportunities in Oldenburg University, Leibniz University and ICTP. I also thank CERN for supporting my visit to its high energy group, as well as ICTP network NT-04 and Sadik Deger for kind supports to attend ``Recent Developments in Supergravity Theories and Related Topics" conference in IMBM, Istanbul. This work has been supported by Riemann Fellowship, and T$\ddot{\text{U}}$BITAK international researchers program 2221. {Finally, I would like to thank the anonymous referee of the paper who emphasized the Lorentz freedom of the BMS constant-time surfaces. }


{\small 
\begin{thebibliography}{10}
\bibitem{BMS} 
  H.~Bondi, M.~G.~J.~van der Burg and A.~W.~K.~Metzner,
``Gravitational waves in general relativity. 7. Waves from axisymmetric isolated systems,''
  Proc.\ Roy.\ Soc.\ Lond.\ A {\bf 269}, 21 (1962).

\bibitem{Sachs}
R.~K.~Sachs,
``Gravitational waves in general relativity. 8. Waves in asymptotically flat space-times,''
  Proc.\ Roy.\ Soc.\ Lond.\ A {\bf 270}, 103 (1962);  ``Asymptotic symmetries in gravitational theory,''
    Phys.\ Rev.\  {\bf 128}, 2851 (1962).

\bibitem{Brown:1986nw} 
  J.~D.~Brown and M.~Henneaux,
  ``Central Charges in the Canonical Realization of Asymptotic Symmetries: An Example from Three-Dimensional Gravity,''
  Commun.\ Math.\ Phys.\  {\bf 104}, 207 (1986).

\bibitem{Barnich:2009se} 
  G.~Barnich and C.~Troessaert,
  ``Symmetries of asymptotically flat 4 dimensional spacetimes at null infinity revisited,''
  Phys.\ Rev.\ Lett.\  {\bf 105}, 111103 (2010),
   \href{http://arxiv.org/abs/0909.2617}{arXiv:0909.2617 [gr-qc]}.

\bibitem{Guica:2008mu} 
  M.~Guica, T.~Hartman, W.~Song and A.~Strominger,
 ``The Kerr/CFT Correspondence,''
  Phys.\ Rev.\ D {\bf 80}, 124008 (2009)
  \href{http://arxiv.org/abs/0809.4266}{arXiv:0809.4266 [hep-th]}.

\bibitem{Carlip:1999cy} 
  S.~Carlip,
  ``Entropy from conformal field theory at Killing horizons,''
  Class.\ Quant.\ Grav.\  {\bf 16}, 3327 (1999)
   \href{http://arxiv.org/abs/gr-qc/9906126}{arXiv:[gr-qc/9906126]}.


\bibitem{Bekenstein:1973ur} 
  J.~D.~Bekenstein,
  ``Black holes and entropy,''
  Phys.\ Rev.\ D {\bf 7}, 2333 (1973).

\bibitem{Hawking:2016msc} 
  S.~W.~Hawking, M.~J.~Perry and A.~Strominger,
  ``Soft Hair on Black Holes,''
  Phys.\ Rev.\ Lett.\  {\bf 116}, no. 23, 231301 (2016)
  \href{http://arxiv.org/abs/1601.00921}{arXiv:1601.00921 [hep-th]}.

\bibitem{Hawking:1976ra} 
  S.~W.~Hawking,
  ``Breakdown of Predictability in Gravitational Collapse,''
  Phys.\ Rev.\ D {\bf 14}, 2460 (1976).

\bibitem{Hawking:2015qqa} 
  S.~W.~Hawking,
  ``The Information Paradox for Black Holes,''
   \href{http://arxiv.org/abs/1509.01147}{arXiv:1509.01147 [hep-th]}.

\bibitem{Mirbabayi:2016axw}
M.~Mirbabayi and M.~Porrati,
``Dressed Hard States and Black Hole Soft Hair,''
Phys. Rev. Lett. \textbf{117} (2016) no.21, 211301
\href{http://arxiv.org/abs/1509.01147}{arXiv:1607.03120 [hep-th]}.

\bibitem{Bousso:2017dny}
R.~Bousso and M.~Porrati,
``Soft Hair as a Soft Wig,''
Class. Quant. Grav. \textbf{34} (2017) no.20, 204001
\href{http://arxiv.org/abs/1706.00436}{arXiv:1706.00436 [hep-th]}.

\bibitem{Bousso:2017rsx}
R.~Bousso and M.~Porrati,
``Observable Supertranslations,''
Phys. Rev. D \textbf{96} (2017) no.8, 086016
\href{http://arxiv.org/abs/1706.09280}{arXiv:1706.09280 [hep-th]}.

\bibitem{Javadinezhad:2018urv}
R.~Javadinezhad, U.~Kol and M.~Porrati,
``Comments on Lorentz Transformations, Dressed Asymptotic States and Hawking Radiation,''
JHEP \textbf{01} (2019), 089
\href{http://arxiv.org/abs/1808.02987}{arXiv:1808.02987 [hep-th]}.

\bibitem{Compere:2019rof}
G.~Comp\`ere, J.~Long and M.~Riegler,
``Invariance of Unruh and Hawking radiation under matter-induced supertranslations,''
JHEP \textbf{05} (2019), 053
\href{http://arxiv.org/abs/1903.01812}{arXiv:1903.01812 [hep-th]}.

\bibitem{Hajian:2015xlp} 
  K.~Hajian and M.~M.~Sheikh-Jabbari,
  ``Solution Phase Space and Conserved Charges: A General Formulation for Charges Associated with Exact Symmetries,''
  Phys.\ Rev.\ D {\bf 93}, no. 4, 044074 (2016)
   \href{http://arxiv.org/abs/1512.05584}{arXiv:1512.05584 [hep-th]}.

\bibitem{Braginsky:1986ia} 
  V.~B.~Braginsky and L.~P.~Grishchuk,
  ``Kinematic Resonance and Memory Effect in Free Mass Gravitational Antennas,''
  Sov.\ Phys.\ JETP {\bf 62}, 427 (1985)
  [Zh.\ Eksp.\ Teor.\ Fiz.\  {\bf 89}, 744 (1985)].

\bibitem{Christodoulou:1991cr} 
  D.~Christodoulou,
``Nonlinear nature of gravitation and gravitational wave experiments,''
  Phys.\ Rev.\ Lett.\  {\bf 67}, 1486 (1991).

\bibitem{Thorne:1992sdb} 
  K.~S.~Thorne,
``Gravitational-wave bursts with memory: The Christodoulou effect,''
  Phys.\ Rev.\ D {\bf 45}, no. 2, 520 (1992).

\bibitem{Strominger:2014pwa} 
  A.~Strominger and A.~Zhiboedov,
  ``Gravitational Memory, BMS Supertranslations and Soft Theorems,''
  JHEP {\bf 1601}, 086 (2016),  \href
  {http://arxiv.org/abs/1411.5745} {arXiv:1411.5745 [hep-th]}.

\bibitem{Strominger:2017zoo} 
  A.~Strominger,
  ``Lectures on the Infrared Structure of Gravity and Gauge Theory,''
 \href{http://arxiv.org/abs/1703.05448}{ arXiv:1703.05448 [hep-th]}.

\end{thebibliography}
}
\end{document}